\documentclass{pasj01}
\usepackage{url}

\def\uncatcodespecials{\def\do##1{\catcode`##1=12}\dospecials}%
{\catcode`\`=\active\gdef`{\relax\lq}}
\def\setupcode 
    {\tt %
     \spaceskip=0pt \xspaceskip=0pt 
     \catcode`\`=\active
     \uncatcodespecials \obeyspaces
     \catcode`\{=1\catcode`\}=2
    }%
\def\SETUPCODE 
    {\tt %
     \spaceskip=0pt \xspaceskip=0pt 
     \catcode`\`=\active
     \uncatcodespecials \obeyspaces
    }%

\begin{document}
\SetRunningHead{Tanaka et al.}{HSC Legacy Archive}

\title{Hyper Suprime-Cam Legacy Archive} 

\author{
  Masayuki Tanaka\altaffilmark{1,2},
  Hiroyuki Ikeda\altaffilmark{3},
  Kazumi Murata\altaffilmark{1},
  Satoshi Takita\altaffilmark{4},
  Sogo Mineo\altaffilmark{1},
  Michitaro Koike\altaffilmark{1},
  Yuki Okura\altaffilmark{1},
  Sumiko Harasawa\altaffilmark{1}
}
\altaffiltext{1}{National Astronomical Observatory of Japan, 2-21-1 Osawa, Mitaka, Tokyo 181-8588, Japan}
\altaffiltext{2}{Department of Astronomy, School of Science, Graduate University for Advanced Studies (SOKENDAI), 2-21-1, Osawa, Mitaka, Tokyo 181-8588, Japan}
\altaffiltext{3}{National Institute of Technology, Wakayama College, 77 Noshima, Nada-cho, Gobo, Wakayama 644-0023, Japan}
\altaffiltext{4}{Institute of Astronomy, University of Tokyo, 2-21-1 Osawa, Mitaka, Tokyo 181-0015,Japan}

\email{masayuki.tanaka@nao.ac.jp}

\KeyWords{astronomical databases: miscellaneous, catalogs, galaxies: general}

\maketitle
\definecolor{gray}{rgb}{0.6, 0.6, 0.6}
\newcommand{\commentblue}[1]{\textcolor{blue} {\textbf{#1}}}
\newcommand{\commentred}[1]{\textcolor{red} {\textbf{#1}}}
\newcommand{\commentgray}[1]{\textcolor{gray} {\textbf{#1}}}


\begin{abstract}
  We present the launch of the Hyper Suprime-Cam Legacy Archive (HSCLA), a public archive of processed,
  science-ready data from Hyper Suprime-Cam (HSC).  
  HSC is an optical wide-field imager installed at the prime focus of
  the Subaru Telescope and has been in operation since 2014. While $\sim1/3$ of the total
  observing time of HSC has been used for the Subaru Strategic Program (SSP), the remainder of the time is used for PI-based programs.
  We have processed the data from these PI-based programs and make the processed, high quality
  data available to the community through HSCLA.
  The current version of HSCLA includes data taken in the first year of science operation, 2014.
  We provide both individual and coadd images as well as photometric catalogs.
  The photometric catalog from the coadd is loaded to the database, which offers a fast access to
  the large catalog.  There are other online tools such as image browser and image cutout tool
  and they will be useful for science analyses.
  The coadd images reach 24-27th magnitudes at $5\sigma$ for point sources and cover approximately 580
  square degrees in at least one filter with 150 million objects in total. We perform extensive quality assurance tests and verify
  the photometric and astrometric quality of the data to be good enough for most scientific explorations.
  However, the data are not without problems and users are referred to the list of known issues
  before exploiting the data for science.
  All the data and documentations can be found at the data release site, \url{https://hscla.mtk.nao.ac.jp/}.
\end{abstract}

\section{Introduction}

The Subaru Telescope has been in operation since 1999.  It has collected an enormous amount of
data from a wide variety of celestial objects and has contributed
to address many of the outstanding astrophysical questions.  Hyper Suprime-Cam (HSC; \cite{miyazaki18}),
a wide-field optical imager installed at the prime focus, was commissioned in 2014 and is now a workhorse instrument.
It has a wide field of view of 1.5 degree diameter, and combined with the light collecting power
of the Subaru telescope, it is the most efficient survey instrument at the time of this writing.
Over the several years of HSC operations, it has made major contributions from solar system bodies
to the most distant galaxies, demonstrating the power of the instrument.

There is an on-going, strategic observing program with HSC (Subaru Strategic Program,
or SSP for short; \cite{aihara18b}), which has been allocated a total of 330 nights.
This program uses a large portion of the total observing time of HSC, and data from
the program have routinely been released to the world-wide community \citep{aihara18a,aihara19}.
The reminder of the HSC observing time is used for PI-based programs, and there is
a large amount of high-quality data from these programs.  The majority of raw data from
these programs are public by now, but the raw data have not been exploited very extensively.
The processing has often been a major hurdle.  This is especially the case for HSC;
it has 104 science CCDs and the pipeline processing requires large computing resources.
This is unfortunate because the scientific value of these data is very high.

In order to improve the situation, we launch the HSC Legacy Archive (HSCLA), where processed,
high-quality HSC data products from PI-based programs are available to the world-wide community.
Efforts similar to HSCLA have been taken at other observatories (e.g., Hubble Legacy Archive;
\cite{jenkner06,whitmore08,whitmore16}) and they are proven to be useful resources
for a wide range of astronomical research. We expect HSCLA to be one of such archives
with large legacy value.

This paper presents the first data release from HSCLA.  The release includes data taken with
HSC as part of PI-based programs in the first year of HSC operation, 2014 (the SSP data are not included).
Both calibrated image and
catalog products are made available to the community.  The images are both individual CCDs and coadds.
The catalogs include positional and shape information of all the detected sources as well as detailed flux measurements.
The measurements come with flags, which indicate possible issues with the measurements.
The coadd catalog is served at a database, providing an easy access to the massive data set.
There are various online data retrieval tools and they also allow users to access the data with no hassle.

The paper is organized as follows.  Section \ref{sec:hscla} introduces the instrument,
followed by a summary of the data included
in this release in Section \ref{sec:data}. Section \ref{sec:qa} presents our quality
assurance tests, and Section \ref{sec:known_issues} summarizes known issues in the data.
Finally, Section \ref{sec:discussion} discusses the future direction of
HSCLA and concludes the paper.  Magnitudes are on the AB system \citep{oke83}.

\section{Hyper Suprime-Cam}
\label{sec:hscla}

\subsection{Instrument}

Hyper Suprime-Cam (HSC; \cite{miyazaki18}) is a gigantic imaging camera for the 8.2 m Subaru telescope built by
National Astronomical Observatory of Japan in collaboration with international partners.
The instrument has 116 detectors, each of which is a highly sensitive 2K$\times$4K deep depletion Hamamatsu CCD \citep{kamata12}.
Eight of them are used for focusing, 4 for guiding, and the remaining 104 for science.
These CCDs are installed inside a vacuum cryogenic dewar operated at -100 $^\circ$C, 
at which the dark current of the CCDs is negligible \citep{komiyama18}.
The science CCDs cover a field of view of 1.7 deg$^2$, which is the largest field of view on
an 8~m class telescope as mentioned earlier.  The instrument saw the first light in the Summer of
2012 and was fully commissioned in 2014.  It has been in science operations since then.
HSC data taken at the telescope are fed to an onsite quick-reduction pipeline to monitor
the sky conditions as well as to check the scientific quality of the data \citep{furusawa18}.
The data are finally transferred to the raw data archive for observers to fetch their data.
Similar to the other instruments on the telescope, the raw data will go public after 1.5 years
of proprietary period.

\subsection{Filter}

\begin{figure}
 \begin{center}
  \includegraphics[width=8cm]{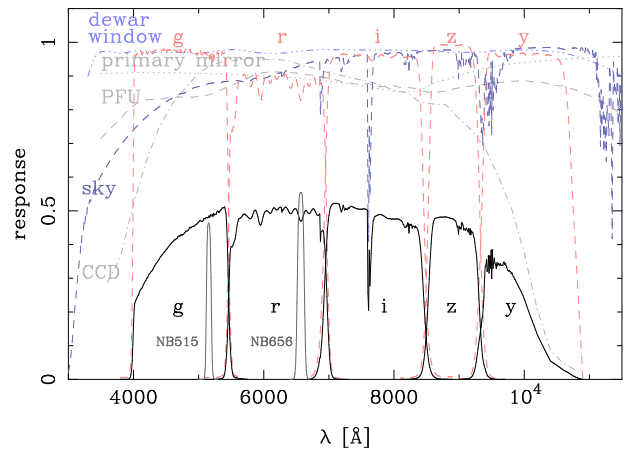}
 \end{center}
 \caption{
   HSC filter responses. The thin lines show various components that contribute
   to the overall response; sky transmission (airmass=1.2 and PVW=2mm), primary mirror of
   the telescope, prime focus optics (PFU), dewar window, filter
   transmission, and CCD response. The solid curves are the total responses.
 }
 \label{fig:filters}
\end{figure}

HSC can house 6 filters at a time. There are 5 broad-band filters ($grizy$) to cover
the wavelength range between 4,000 $\rm\AA$ to 10,000$\rm\AA$. In addition, there are
several narrow-band filters primarily designed to detect emission line objects.
\citet{kawanomoto18} summarizes the HSC filter system.  All the broad-band filters as well as
two narrow-band filters, NB515 and NB656, are included in this release, although the field
coverage of the narrow-band filters is fairly small.  Fig.~\ref{fig:filters} summarizes
these filters.

As discussed in \citet{kawanomoto18}, the $r$ and $i$-band filters have a significant
spatial variation in their transmission curves.  It is mostly a function of distance
from the filter center, and the telescope dithering averages the variation to some extent.
However, in order to reduce the problem, the new filters, $r_2$ and $i_2$, with much weaker
radial dependence have been manufactured and been in use since 2016.  The current release
includes only data from 2014 and thus all the $r$ and $i$-band data are taken with the old filters.

\section{Data}
\label{sec:data}

\begin{figure*}
 \begin{center}
  \includegraphics[width=16cm]{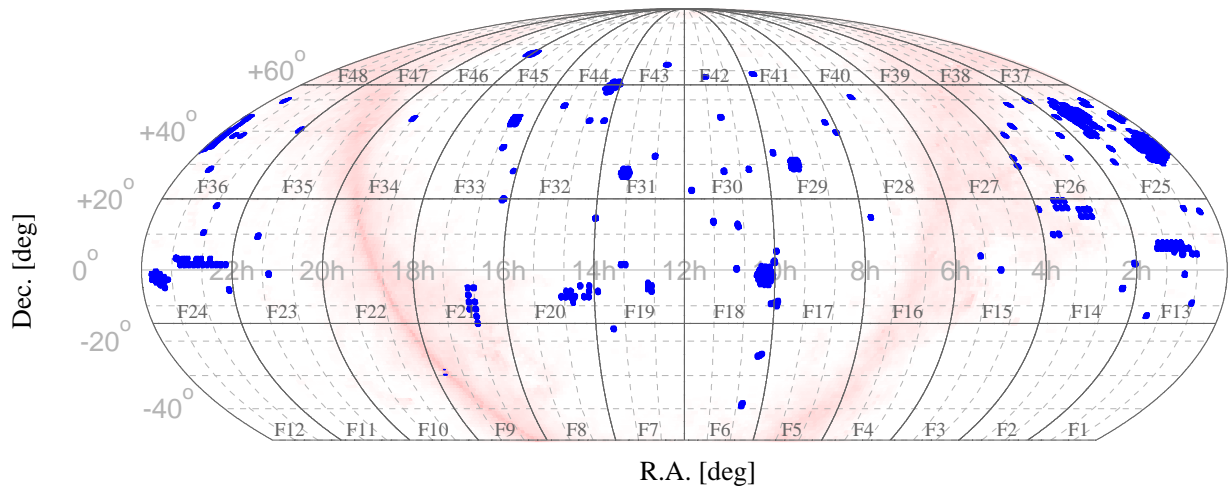}
 \end{center}
 \caption{
   Field coverage.  The blue regions show the area covered in at least one filter.
   We define 48 regions of the sky observable from Subaru shown as F1 to F48.
   Quality assurance tests in Section \ref{sec:qa} are performed in each field separately
   for illustrative purposes.  The pink shades in the background is the Galactic extinction
   map from \citet{schlegel98}.
 }
 \label{fig:fields}
\end{figure*}

Fig.\ref{fig:fields} shows the field coverage of the data in this release.
The spatial coverage is quite patchy because the data are a collection of PI-based programs with
a wide variety of science goals.  For instance, the relatively large coverage in F25 is from a campaign
to observe the outskirts of the Andromeda galaxy.  Although the data are taken for various purposes,
they are processed in the same manner.  This section summarizes the data screening and processing.

Before we describe the processing in detail, it is important to define the terminology used in the processing.
{\tt Visit} is an integer identifier uniquely assigned to an exposure.  It is always an even number
incremented by 2 for each exposure.  In the first phase of the pipeline processing, each visit is processed
separately to remove instrumental signatures and also to perform initial photometric and astrometric
calibrations.  Later stages include multi-visit processing and are performed separately in each of
a pre-defined, rectangular region of the sky.  This region is called {\tt tract} and is about 1.7 degree on a side.
There is a small overlap between the adjacent tracts ($\sim1'$ on the equator).  A tract is further subdivided
into $9\times9$ regions, each of which is called {\tt patch}.  A patch is 4200 pixel or $\sim12$ arcmin on a side,
and there is again an overlap between neighboring patches (100 pixels).  The combination of tract and patch is
a unique identifier of a sky region and is frequently used at the database described below.

\subsection{Data Screening}

For the initial release of HSCLA, we focus on data from 2014, which is the first year of HSC operation.
After excluding all test/engineering visits as well as SSP visits, we process 8 CCDs (4 around
the field center and the other 4 on the edge) from the remaining visits to measure the sky conditions.
This is an important step because not all the visits are taken under good photometric conditions
and a careful screening of the visits is required.
We choose to use visits with sky transparency better than 50\% and seeing below 1.3 arcsec.
As the $g$-band visits tend to have worse seeing than the other filters, we relax
the seeing constraint to 1.5 arcsec only for the $g$-band.
Note that we do not apply any selection on the nature of the targets.

\subsection{Pipeline Software and Data Processing}

The data processing pipeline is a version of the LSST pipeline \citep{juric17,ivezic19} with
HSC-specific features included and is called {\tt hscPipe}.  In this subsection, we briefly describe the pipeline with
an emphasis on differences from the standard processing manner as well as on features that
users should be aware of.  The algorithmic detail of the pipeline can be found in \citet{bosch18}
with updates described in \citet{aihara19}.

It is important to note that we use {\tt hscPipe v6.7}, which is the same version as used in
the public data release 2 (PDR2) from HSC-SSP \citep{aihara19}.  All the configuration parameters
are the same as well.  PDR2 has been used widely
and been demonstrated to be of high scientific quality.   The characteristics of the processed data
from HSCLA should be very similar and we encourage readers to refer to the PDR2 paper \citep{aihara19}
as well as to the data release site (\url{https://hsc-release.mtk.nao.ac.jp}).  In particular,
the list of known issues there are fully relevant to this release of HSCLA.

After the screening of the data,
the pipeline starts with detrending individual CCDs; overscan subtraction, bias and dark subtraction,
flat-fielding, bad-pixel masking, linearity correction, cross-talk correction, and the correction
for the brighter-fatter effect \citep{antilogus14}.
The local sky background is then subtracted.
Only in the $y$-band, the fringe removal and the scattered encoder light removal are
performed prior to the background subtraction \citep{aihara19}.
Sources are detected for PSF determination, photometric and astrometric calibrations.  
The photometry and astrometry are calibrated against Pan-STARRS 1 (PS1) $3\pi$ catalog
\citep{schlafly12,tonry12,magnier13,chambers16}, whose astrometry is further calibrated
against Gaia DR1 \citep{gaia16a,gaia16b}.

The CCD processing is successful for most visits, but there are some failed visits, most of which
targeted very nearby extended galaxies.  Currently, the pipeline is not optimized for such objects,
and the processing fails when a majority of the CCD area is occupied by an object.
As we discuss in Section \ref{sec:known_issues}, this is a typical failure mode in the current release.
There are other issues that users should be aware of and they are summarized in Section \ref{sec:known_issues} as well.
The numbers of visits successfully processed are 1535, 427, 1582, 193, 151, 6, 36 for $g,\ r,\ i,\ z,\ y$,
NB515 and NB656, respectively.

We then use common stars from multiple visits to solve for photometry and astrometry to achieve
better accuracy.  After that, the local sky background subtracted in the CCD processing is put back in,
and the global sky subtraction algorithm \citep{aihara19} is applied to subtract the sky, while
preserving wings of bright objects.  The CCD images are then warped into a patch and coadded together.
Note that separate observing programs may have targeted the same (or overlapping) fields, but they are
coadded together to gain the depth.
Sources are detected on the coadd image in each band separately.  The source lists from available
filters are merged into a master source list.  Based on this list, measurements are performed in
two steps.  First, measurements are performed in each band separately allowing the object centroids
and shapes to vary in each band. From these measurements, we choose one reference band for each object.
This is the $i$-band in most cases, but other filters can be chosen when an object has a low $S/N$
in the $i-$band.  Finally, we perform 'forced' measurements by applying the object centroids and
shapes from the reference filter to all the other filters.  The first measurements
are stored in the {\tt meas} talbes and the second measurements in the {\tt forced} tables in
the database discussed below.

\subsection{Data Products}

All the pipeline products, including intermediate files, are made available at HSCLA.
The pipeline products include processed CCD images, source catalogs from the CCDs, images warped to patches,
coadd images, and source catalogs from the coadds, among others.  All these images are calibrated photometrically
and astrometrically and the catalogs include source positions, shapes (2nd order moments), and various flux measurements.
The total area observed in at least one filter amounts to 580 square degrees.  Each filter covers 320, 48, 470, 38, 13,
2, and 2 square degrees in $g,\ r,\ i,\ z,\ y,$ NB515, and NB656, respectively.  The number of objects detected in each band is
83, 15, 93, 8, 3, 0.4, 0.2 million objects respectively in these filters.
The depth varies significantly across the field and in different filters but is generally in the range of
24-27th magnitude.  See further discussion in Section \ref{sec:qa}.

There are several popular flux measurement techniques.  One of them is the PSF photometry, which is matched-filter
photometry using the PSF model and is expected to deliver the best photometry for point sources (it is less useful for
extended sources).  For extended sources, one can use
{\tt CModel}, which is a model fit with the exponential and de Vaucouleurs profiles combined taking into account the PSF
in each band.  It asymptotically approaches the PSF photometry for point sources. Another popular measurement is
the Kron photometry.  There is also PSF-matched aperture photometry with various target PSF sizes.  As we discuss in
Section \ref{sec:known_issues}, the PSF-matched photometry can be quite useful in some cases.
Note that the catalogs contain objects in the intermediate steps of deblending and it is important to select
objects after the deblending only for scientific analyses.  Note as well that objects in the overlapping regions of adjacent
patches and tracts are measured multiple times and one needs to eliminate the duplication.  The easiest way to do both
is to use the {\tt isprimary} flag; the flag is set for deblended objects (including isolated objects for which the deblender
does not run) located in the inner regions of tract and patch.  The flag is called in slightly different ways
between the fits file and database, but it should be easily identifiable.

A crude star-galaxy separation is performed based on the magnitude difference between the PSF and CModel magnitudes
(the difference is small for point sources).   The {\tt extendedness} parameter is the result of  this measurement.
It works well down to $\sim23$rd magnitudes, but the completeness and contamination depends strongly on the exposure
time and seeing.  See the HSC-SSP PDR2 site for a discussion (\url{https://hsc-release.mtk.nao.ac.jp/}).

The coadd catalogs are loaded to the database.  As each object has many measurements, the database tables are split
into multiple smaller tables.  There is a summary table, in which most popular and useful columns are carefully selected
and combined.  Most users should start with this summary table.  In the current version of HSCLA, individual CCD catalogs
are not loaded and one needs to look at fits files for time-series analyses.

There are a few additional products generated separately from the pipeline and are stored in the database.
One is the bright star mask.  There are many bright Galactic stars in the HSCLA region and
each star has a bright halo, bleeding trail and other features.  Because the sky background is subtracted on
a large scale, the wings of bright stars are preserved well.  Due to the elevated background,
a large number of fake sources are often detected in the halo, although the pipeline does attempt to reduce such spurious sources.
An effective workaround is to mask regions around bright
stars.  The number of sources as a function of distance from the center of stars is used as an indicator of
effects of stellar halo and other features and the mask sizes are determined empirically from the number of detected sources.
The mask size is dependent on the brightness of a star and we use the Gaia $B_p$ and $R_p$ photometry to infer magnitudes
of stars in the HSC filter system.  There is an online documentation about how the mask sizes are determined.

The other is a set of random points drawn for each patch and for each filter with a density of 100 points
per square arcmin.  These random points come with flags propagated from the coadd images (saturated, interpolated, etc)
as well as with useful information such as the number of coadds contributed to the point, pixel variance, sky background,
and PSF size.  The randoms can be used for several purposes such as a random sample for clustering analysis,
identifying problematic area, computing the survey area, possible issues with the background, etc, and are
a quite useful product.

\section{Quality Assurance Tests}
\label{sec:qa}

Following the data processing, we now check the astrometric and photometric quality of
the data.  We carry out a number of tests in all the fields shown in Fig.~\ref{fig:fields},
but we present some of the most important tests and show only 1 field for each test in this Section.
The whole suite of the quality assurance figures can be found at the data release site.

\begin{figure*}[ht]
 \begin{center}
  \includegraphics[width=12cm]{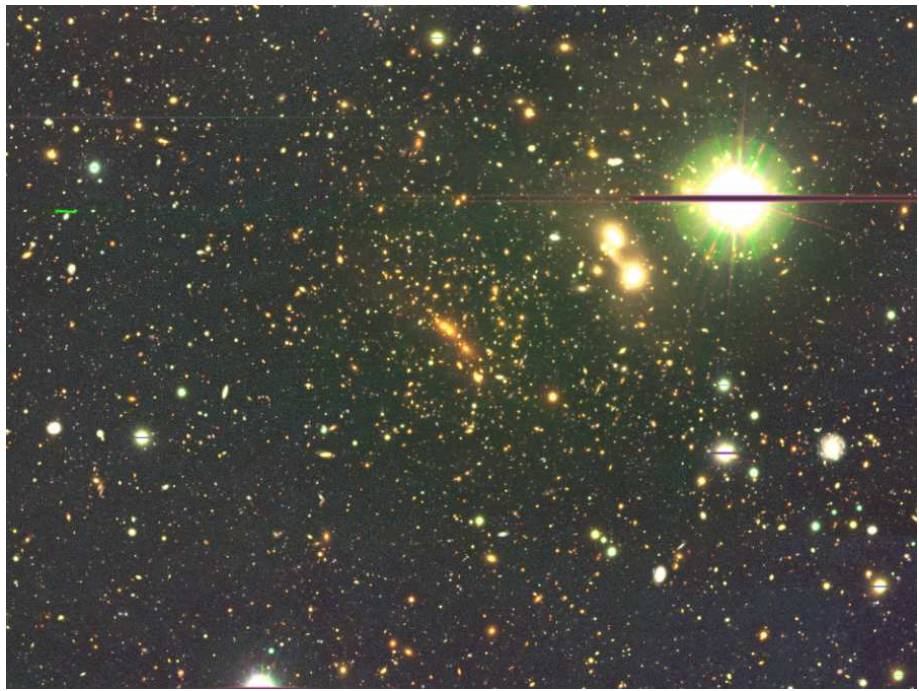}
 \end{center}
 \caption{
   A blow-up of the central part of the CL0016+16 galaxy cluster.  This color image is based on the $gri$ images.
 }
 \label{fig:cl0016}
\end{figure*}

Most regions are covered in 2 filters and only a small fraction of the observed area
in this release is covered in more than 2 filters.  This will improve in our future releases;
we will include more data and more filters over a wider area.
Fig~\ref{fig:cl0016} shows a color image of the CL0016+16 cluster at $z=0.55$, which is observed
in all the 5 broad-band filters.  There is prominent large-scale structure around the cluster
extending over a degree scale \citep{tanaka09}. It is impressive that the whole structure
discussed in \citet{tanaka09} is covered in a single pointing of HSC.

\subsection{Astrometry}
\label{sec:qa_astrometry}

We first focus on astrometry.  We select stars brighter than the 20th magnitude in each filter
from HSCLA and match them against Gaia DR1 \citep{gaia16a,gaia16b}.
We use the extendedness parameter to select stars. At this bright magnitude range, the contamination
of galaxies is negligible.
There is a newer release from Gaia, but we have used DR1 following HSC-PDR2.  We plan to use a newer version in our future releases.
We exclude saturated stars from the matched catalog using the saturation flag.  The saturation limit depends on the seeing and
exposure time, but most of the stars used here are fainter than the 18th magnitude.

\begin{figure*}[ht]
 \begin{center}
  \includegraphics[width=16cm]{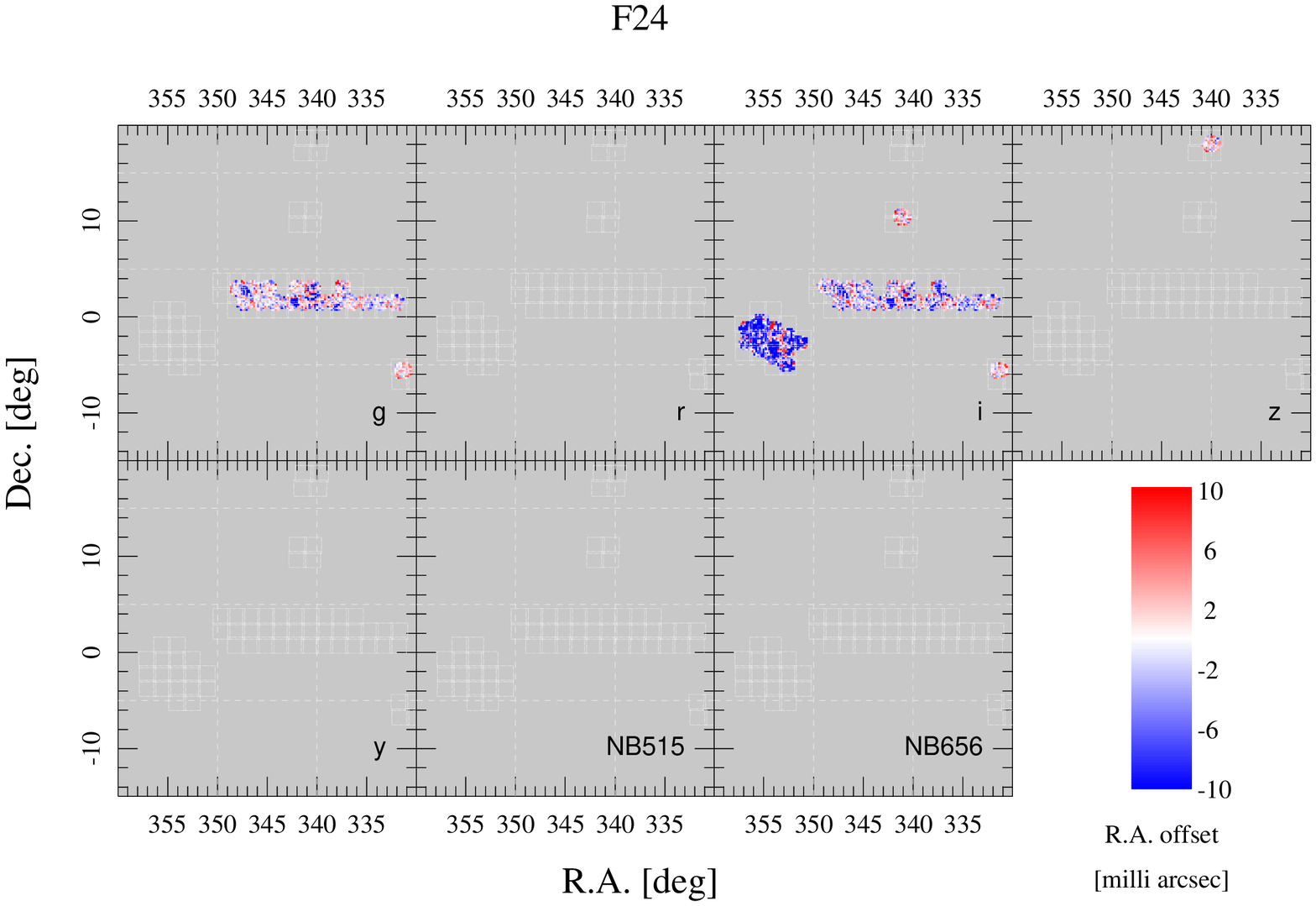}
 \end{center}
 \caption{
   Astrometric offsets in R.A. with respect to the Gaia catalog for point sources.
   The panels show different filters with the filter name indicated in each panel.
   The dashed lines show the tract borders.  The mean offset is computed for reach patch
   and is color-coded as shown in the figure.
   This figure is for one of the fields defined
   in Fig.~\ref{fig:fields} and figures for all the other fields can be found at the data release site.
 }
 \label{fig:ra_offset}
\end{figure*}

Fig.~\ref{fig:ra_offset} shows the astrometric offset in the direction of R.A. with respect to Gaia.
As our astrometric calibrations are performed against Gaia\footnote{
To be precise, we calibrate against the PS1 catalog, whose astrometry is calibrated against Gaia.
}, this is not an external comparison,
but it is still a useful check for the calibration accuracy. The mean offset is
computed for each patch and the color-coding is also for each patch.  The number of stars used
varies significantly from patch to patch, but it is typically 50.  In this particular example,
there are data only in $giz$ filters and the panels for the other filters are left blank.
As can be seen, the astrometry is overall fine, although there are small-scale structures with
a scale of several patches that show $>10$ milli arcsecond offsets.  There is also a larger scale
structure observed in the $i$-band (the chunk on the left shows a systematic offset).
Some of these offsets might be due to proper motion of stars, which is not accounted in the astrometric
calibration, but further investigations are needed here.
We find that only about 0.5\% of the patches have offsets larger than 50 milli arcseconds, which
suggests that the overall astrometric accuracy is good.  Note that the offsets are averaged over
a patch and there may be individual sources that have a larger offset.  The corresponding figures for
the Dec. direction are available at the data release site.

\subsection{Photometry}
\label{sec:qa_photometry}

\begin{figure*}
 \begin{center}
  \includegraphics[width=16cm]{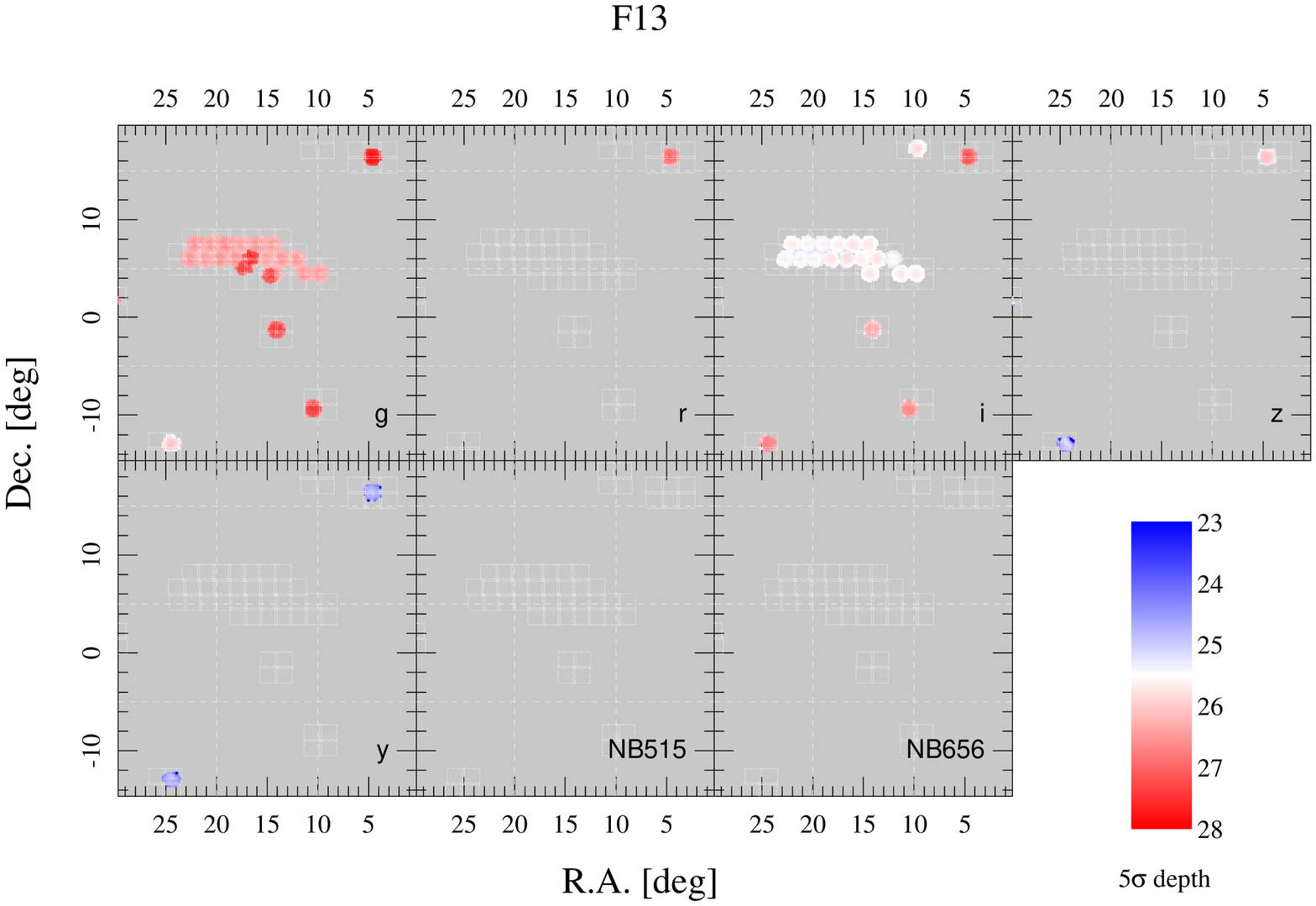}
 \end{center}
 \caption{
   Same as Fig.~\ref{fig:depth}, but for $5\sigma$ limiting magnitudes for point sources for F13.
 }
 \label{fig:depth}
\end{figure*}

\begin{figure*}
 \begin{center}
  \includegraphics[width=16cm]{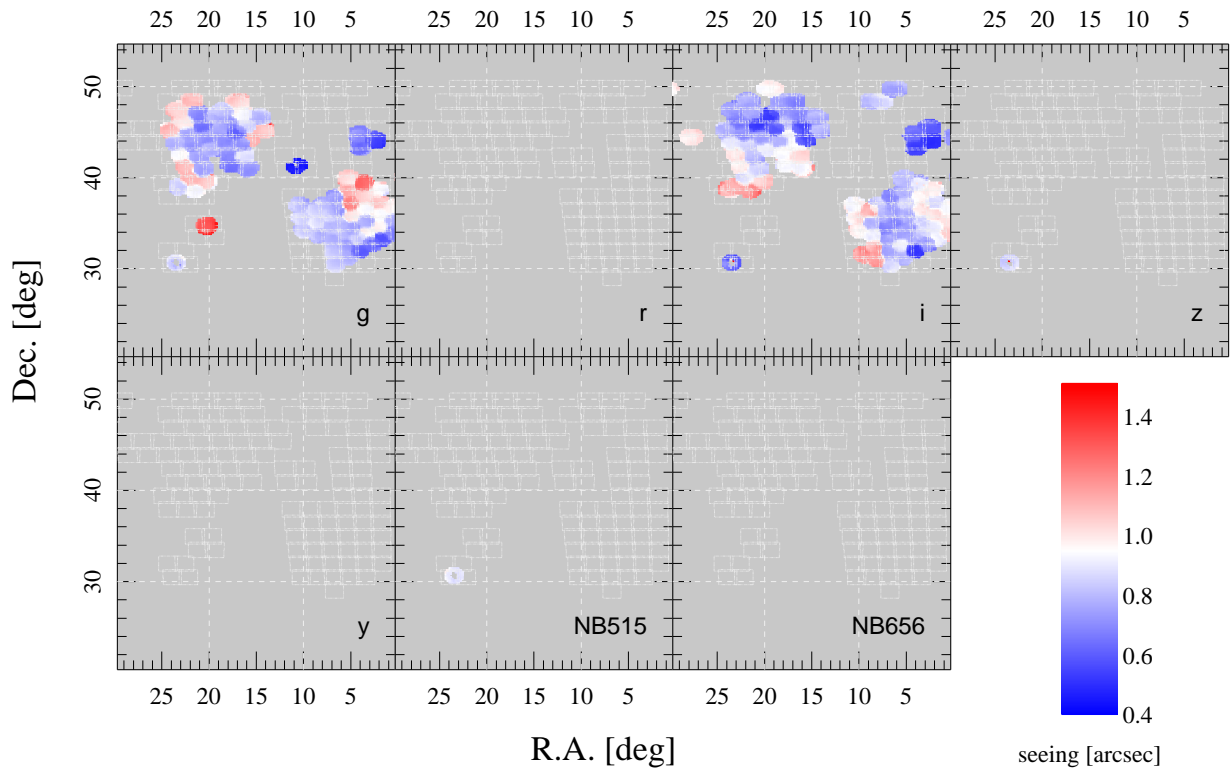}
 \end{center}
 \caption{
   Same as Fig.~\ref{fig:depth} but for seeing FWHM in arcsec in F25.
 }
 \label{fig:seeing}
\end{figure*}

We move on to discuss photometry.  To give an overview of the data quality, we show in
Figs.~\ref{fig:depth} and \ref{fig:seeing} the spatial distributions of the $5\sigma$ limiting
magnitudes for point sources and seeing FWHM, respectively.
The $5\sigma$ limiting magnitudes
for point sources are from the forced catalog and are estimated as magnitude of stars with $S/N_{PSF\ flux}=5$
using the photometric uncertainty quoted from the pipeline.
The seeing is measured from the 2nd-order moment of bright stars ($<21$~mag) and translated into the Gaussian FWHM.
The depth is generally shallower in redder bands and there is a significant spatial variation as expected.
The seeing also varies significantly over the fields; from 0.4 arcsec to 1.5 arcsec.
Users are encouraged to first refer to the depth and seeing map to see the data quality
in their fields of interest.

\begin{figure*}
 \begin{center}
  \includegraphics[width=16cm]{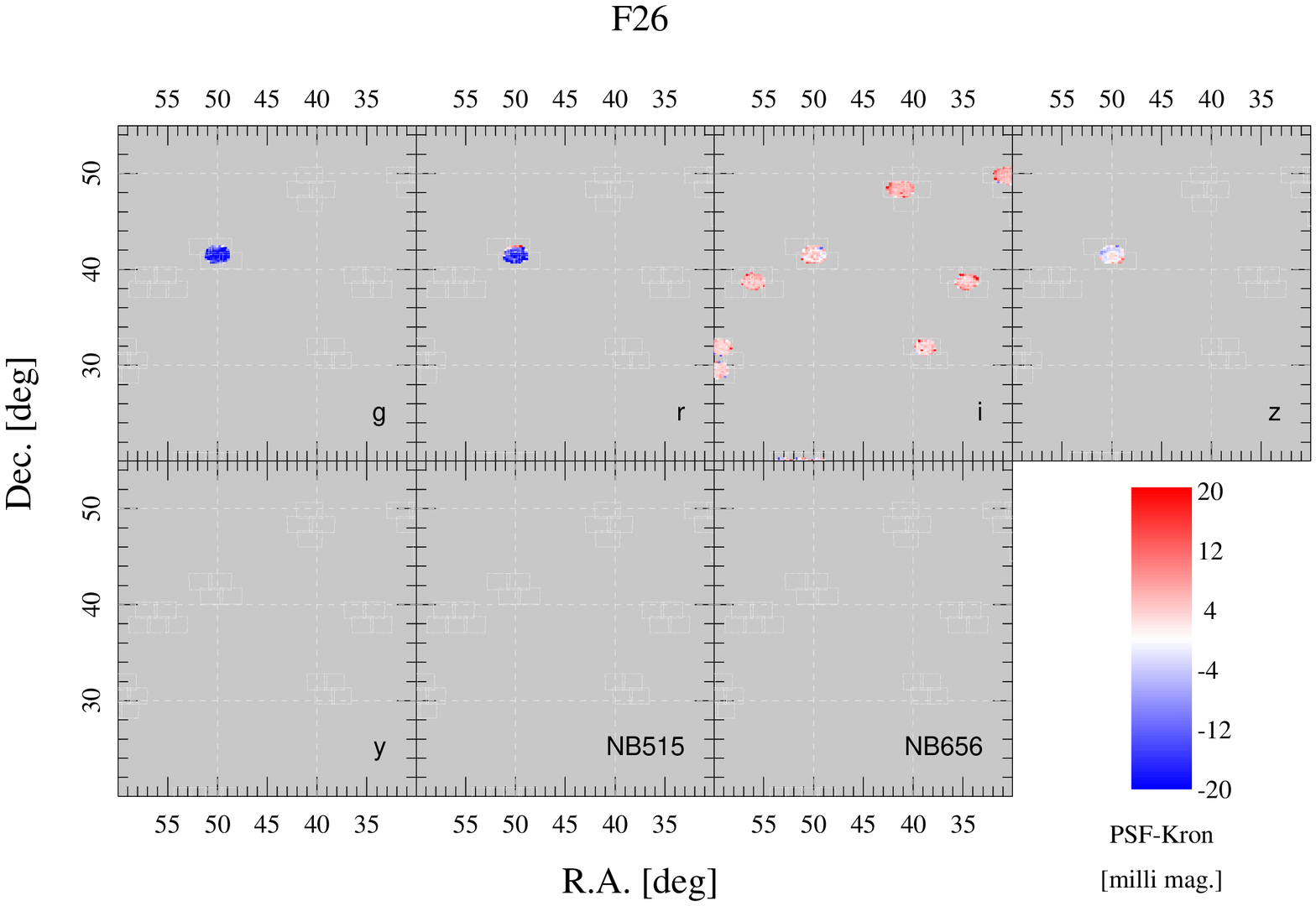}
 \end{center}
 \caption{
   Same as Fig.~\ref{fig:depth}, but the color scale shows the difference between PSF magnitude and
   Kron magnitude for point sources.
 }
 \label{fig:psfkron}
\end{figure*}

Next, we perform internal consistency checks of our photometry.  There are a few ways to
check the internal consistency, but we here test the convergence of Kron and CModel photometry
to the PSF photometry for point sources.  Fig.~\ref{fig:psfkron} shows the difference between
the PSF and Kron magnitudes for point sources.  For the internal consistency checks here, we use
point-like objects down to the 22nd mag in each filter.  Again, the saturated sources are excluded
and the contamination of galaxies is negligible at this magnitude range.  The magnitude offsets are averaged over
a patch as done before.  Most regions have magnitude difference less than 10 mmags.
Although there are regions with larger differences, we find that only 3\% of the patches in this release have
offsets larger than 20 mmags.  This shows a good internal consistency between Kron and PSF magnitudes
for point sources.  Figures for the CModel photometry can be found online.

\begin{figure*}
 \begin{center}
  \includegraphics[width=16cm]{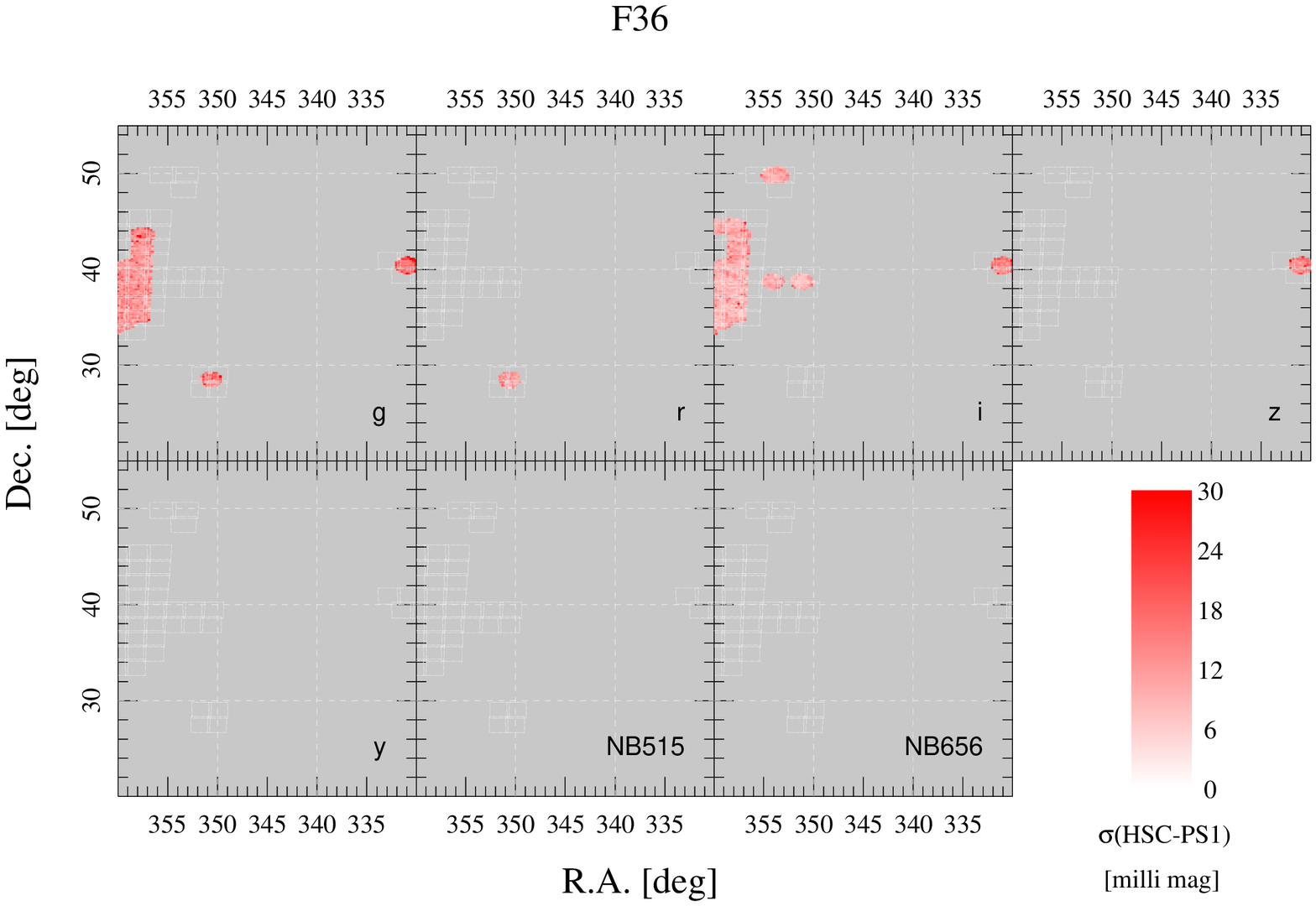}
 \end{center}
 \caption{
   Same as Fig.~\ref{fig:depth}, but for magnitude difference between HSC and PS1 in F26.
 }
 \label{fig:psfcal}
\end{figure*}

Finally, we compare PSF photometry of bright ($<20$~mag) point sources against photometry from PS1.
As we calibrate our photometry against PS1, this is not an external comparison, but it is
still a useful calibration check.  Fig.\ref{fig:psfcal} shows this comparison.
On average, the agreement is fairly good with a scatter of 10~mmag.  Note that the photometric
uncertainty of PS1 is subtracted from the quadrature here.  If we look at the entire area in this release,
the mean offset is 4 mmags with a dispersion of 5 mmags.  Again, the mean is computed over a patch
and individual objects will show a larger dispersion, but the overall accuracy is fairly good.

\begin{figure}[ht]
 \begin{center}
  \includegraphics[width=8cm]{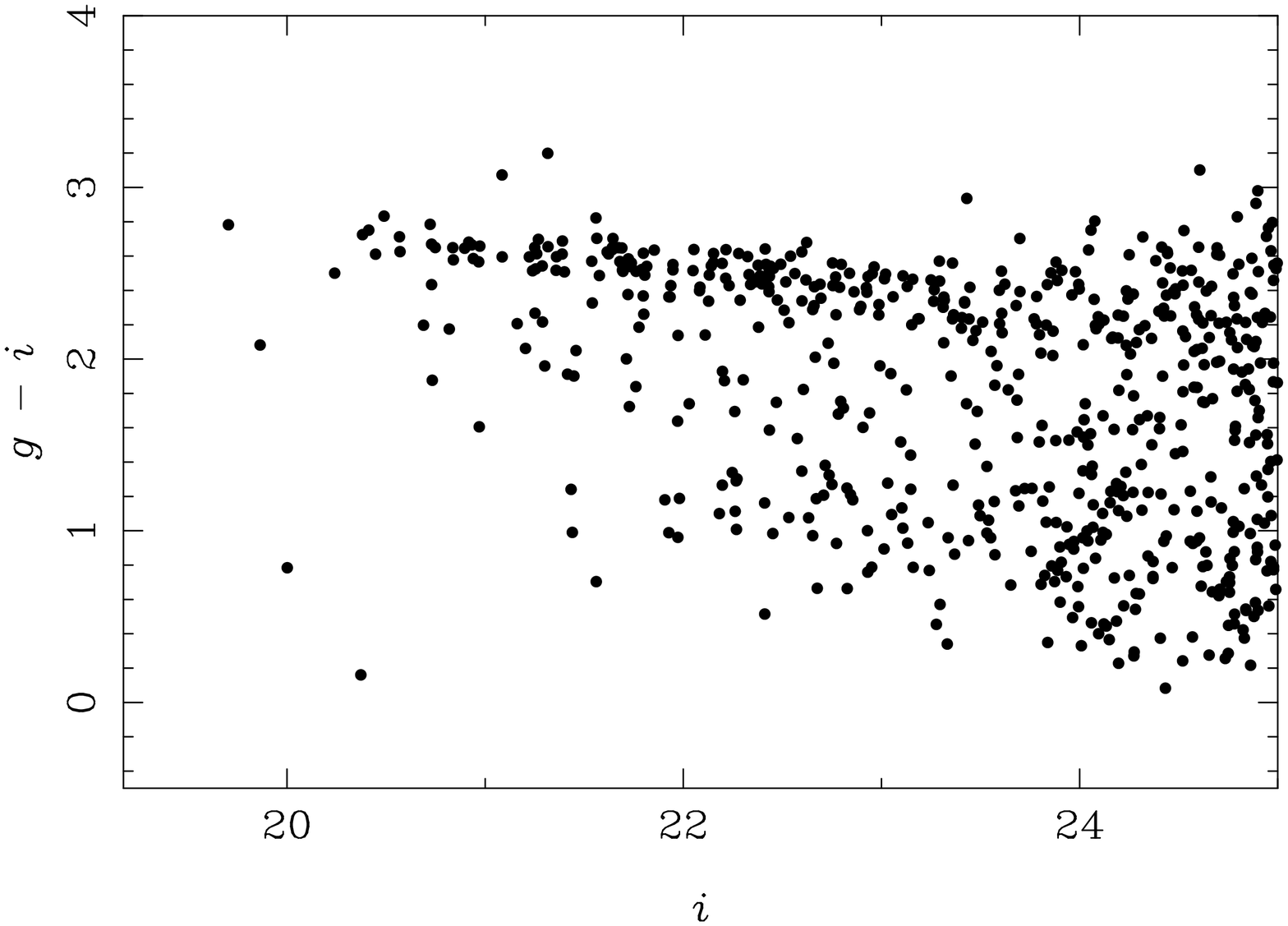}
 \end{center}
 \caption{
   $g-i$ plotted against $i$ for galaxies in the core of the CL0016 cluster.
 }
 \label{fig:cl0016_cmd}
\end{figure}

These tests suggest that the astrometric and photometric accuracy of the data is sufficiently good
for scientific explorations.
Fig. \ref{fig:cl0016_cmd} shows the color-magnitude diagram of galaxies in the core of the CL0016 cluster
shown in Fig. \ref{fig:cl0016}.  This is a very massive cluster and the red sequence is prominent
down to $i\sim25$.  Here, we use the PSF-matched aperture photometry following the discussion in the next Section,
but the red sequence is also clear if we use CModel.  This figure further illustrates the quality of HSCLA.
However, the data are not without problems and we will summarize the known
problems in the next section.  Once again, the data quality (and problems, too) should be identical
to that of HSC-PDR2 and we encourage readers to refer to \citet{aihara19}.

\section{Known Issues}
\label{sec:known_issues}

As discussed in the previous section, the data are in general of science quality, but
there are some processing failures and features that need to be taken care of.
This section summarizes these known issues.  The issue list is also available at
the data release site and we encourage users to refer to it before using the data for science.
We will keep the online list up to date.

\subsection{Remaining artifacts}
While we make an attempt to remove artifacts such as satellite trails and cosmic rays in the processing,
some of them make it to the coadd.  This is combination of reasons; only a small number
of visits are taken, which makes it difficult to identify artifacts in the current algorithm \citep{aihara19},
dither lengths are not sufficiently large and optical ghosts stay at the same position, etc.
We hope to achieve better removal of some of these artifacts with a future version of the pipeline,
but issues with the data themselves (e.g., dither lengths being too short) are something
we can not change.  There is an online image browser (hscMap) and it is a good practice to visually
check the images of the field of interest for any obvious artifacts, which may contaminate
science analyses.

\subsection{Processing failures around very large galaxies}
\label{sec:processing_failures}
Because of the large field of view of HSC, some of the observing programs aimed at
very nearby, significantly extended galaxies.   The CCD processing often fails in
the central part of these objects because a CCD is filled with objects and the pipeline
cannot solve for astrometry and photometry.  In addition, the global sky subtraction
is not 'global' enough, and we tend to subtract parts of the galaxies as background.
These failures lead to a lack of coadd images in the central parts and fluctuating
background in the outskirts around the nearby galaxies.
For instance, M33 
is observed in $giz$ and $NB515$ and the processing failed badly in the $i$-band.
The other bands may look fine, but a careful inspection reveals structure such as CCD edges.
Also, the scaling of the image variance applied in the coaddition phase, which is meant to
deliver more accurate photometric uncertainties, is unusually large, which also indicates
a processing failure.  As the current pipeline is not optimized for very extended sources,
a care has to be taken when analyzing them.

\subsection{Deblending failures}
As discussed in \citet{aihara18a}, the deblender may fail in crowded regions such as
the core of a galaxy cluster.  A workaround is to use PSF-matched aperture photometry on pre-deblend
images ({\tt undeblended\_convolvedflux} in the database) as shown in Fig.~\ref{fig:cl0016_cmd}.
A set of target seeing sizes to match the PSF to and aperture sizes to do photometry is available.
As this measurement is not performed when the original seeing is worse than target seeing, users may want
to check the seeing sizes of a field of interest and use a large enough target seeing size.
Note that this workaround may not work in extremely crowded regions such as star clusters or the Galactic bulge.
The crowded field photometry is a different technique from the standard deblending algorithm and such
a functionality is not implemented in the pipeline yet.  Users will have to run an external code such
as {\tt DAOPHOT} \citep{stetson87} on the HSC images.

\subsection{Extended faint sources}
There are sources that are extended and have low CModel fluxes.  Visual inspections of
them reveal that most of them are not real sources and are often located in the outskirts
of large objects.
We observe similar issues in HSC-SSP PDR2 \citep{aihara19} and these objects can
be a significant source of contamination for faint object research.
Investigations are under way, but {\tt undeblended\_convolvedflux} could be used instead of CModel;
CModel is sensitive to outer parts of galaxies, while the aperture photometry is not.
We hope to provide a workaround for these extended sources in the future.

\subsection{Possible background residual}
The global sky subtraction scheme we apply preserves the wings of bright objects well.
A possible side effect is that a small-scale background fluctuations are left in the images.
This may be contributing to the extended faint sources mentioned above.  The level of
this possible residual sky is fairly low ($\sim29\rm~mag\ arcsec^{-2}$) and most sources
are not significantly affected.  To estimate the background level at the catalog level,
one can use 'sky objects', which have measurements performed on the blank sky.
They can be identified as {\tt merge\_peak\_sky=True} at the database.

\subsection{Problematic coadds}
Some of the coadd images have issues such as bad background subtraction
(as discussed in Section \ref{sec:processing_failures}) and scattered light
with significant spatial structures.  They are not sufficiently big problems to cause
processing failures, but the coadds and resulting photometric catalogs should be used with caution.
We compile a list of such problematic tracts in Table \ref{tab:problematic_tracts}.

\begin{table*}[htbp]
  \begin{center}
    \begin{tabular}{lll}
      \hline
      Tract       &  Filter     &  Comments\\
      \hline
      14135 (M33) & $g,i,NB515$ & Variance scaling applied in the pipeline too large\\
      14134,14135,14345,14346 (M33) & $i$ & Poor sky subtraction\\
      15528,15529,15713,15714 (M31) & $g$ & Poor sky subtraction\\
      6995,6996,7230,7231,7467,7468,7705,7706,7947,7948,8187 & $r$ & Scattered light\\
      16031,16032,16205,16206 & $i$ & Poor sky subtraction (scattered light?)\\
      9221,9222,9946,9947,10188,10189 & $i$ & Tracts overlapping with bad data mistakenly processed\\
      7083,7319,7320 & $z,y$ & Poor sky subtraction (scattered light?)\\
      \hline 
    \end{tabular}
  \end{center}
  \caption{
    Problematic tracts.
  }
  \label{tab:problematic_tracts}
\end{table*}

\section{Summary and Future Prospects}
\label{sec:discussion}

We have presented the first release of HSCLA, which includes data from the PI-based programs in the first year
of the HSC science operation, 2014.  The data from HSC-SSP is not included here.  We have retrieved data taken
under reasonable photometric conditions and processed them with {\tt hscPipe v6.7}, which is the same version of
the pipeline used for HSC-SSP PDR2 \citep{aihara19}.  The characteristics of the processed data should thus be similar.
As demonstrated in Section \ref{sec:qa}, we have achieved good photometric and astrometric accuracy
and the data area ready for a wide range of scientific explorations.  We make a total of 150 million
objects over 580 square degrees available to the community in this release.

The data are served at \url{https://hscla.mtk.nao.ac.jp/}, in a similar fashion to HSC PDR1 and 2.
A full set of quality assurance figures are available there and users are referred to them before making
use of the data products for science.  The online list of known issues is kept up-to-date and it should
be referred to as well.

All the files from the pipeline, including intermediate data products, are available.  Most of them
are in the fits format and users can either follow the directory tree or use an online file search tool
to look for files of interest.  {\tt hscMap}, an online image browser, is a very efficient tool to navigate
yourself around HSCLA and see what data are available.  Users can choose which filters to use to create
color images, and also zoom in/out and pan to look at the images.  By selecting a region inside {\tt hscMap},
users can query the database for that region.  It also accepts a user catalog.
There is an image cutout tool, with which users can retrieve image cutouts,
as well as a tool to retrieve PSF models at objects' positions.   These online tools come with user manuals
and readers are refereed to them for detailed usage.

The coadd catalogs are loaded to the database.  Both meas and forced catalogs have been loaded and they are
split into multiple smaller tables.  It is somewhat cumbersome to join these tables and queries may be slow.
We thus offer a summary table, where most important columns are carefully selected and combined into a single table.
There are two versions of the summary table: one has photometry in fluxes, and the other has magnitudes.
We suggest users start with the summary tables.
The schema browser should be referred to for a summary of the database contents.
The catalogs from the individual CCD images are not loaded at this point.
We ask users to acknowledge us when they publish their results based on data from HSCLA.
The standard acknowledgment text is provided at the data release site.

Looking towards the future, the database currently offers data download only, but we plan to
extend the system to a Jupyterhub-like system, where users can perform scientific analyses next to the data
without a massive transfer. This will be a major upgrade to HSCLA.
Also, there are a plenty more public
data in the raw data archive.  As mentioned earlier, all data taken at the Subaru Telescope are made available
after a nominal proprietary period of 1.5 years.  As of this writing, data up through mid-2019 are available.  We plan to process
data taken after 2014 and make further releases in the future.  Once HSC-SSP is over, raw data
from HSC-SSP will also be entirely public, and we plan to merge all the SSP data to HSCLA,
so that HSCLA will be the central data repository for all the processed HSC data.  We continue to put efforts in
HSCLA to increase the legacy value of the archive, so that the community can fully exploit the scientific value of
the HSC data and address important, outstanding questions in astrophysics.

\section*{Acknowledgments}

This paper is based on data collected at Subaru Telescope, which is operated by the National Astronomical Observatory of Japan.  
We thank all the Subaru Telescope staff for the continued work to operate the telescope and instrument.
This work would not have been possible without their efforts over many years.  We also thank Hisanori Furusawa for his continued support to HSCLA.

This paper makes use of software developed for the Large Synoptic Survey Telescope. We thank the LSST Project for making their code available as free software at http://dm.lsst.org

The Pan-STARRS1 Surveys (PS1) and the PS1 public science archive have been made possible through contributions by the Institute for Astronomy, the University of Hawaii, the Pan-STARRS Project Office, the Max Planck Society and its participating institutes, the Max Planck Institute for Astronomy, Heidelberg, and the Max Planck Institute for Extraterrestrial Physics, Garching, The Johns Hopkins University, Durham University, the University of Edinburgh, the Queen's University Belfast, the Harvard-Smithsonian Center for Astrophysics, the Las Cumbres Observatory Global Telescope Network Incorporated, the National Central University of Taiwan, the Space Telescope Science Institute, the National Aeronautics and Space Administration under grant No. NNX08AR22G issued through the Planetary Science Division of the NASA Science Mission Directorate, the National Science Foundation grant No. AST-1238877, the University of Maryland, Eotvos Lorand University (ELTE), the Los Alamos National Laboratory, and the Gordon and Betty Moore Foundation.

\bibliographystyle{apj}
\bibliography{references}

\end{document}